 \documentstyle[12pt] {article}
 
\begin{document}
 \title{ {\small { \bf SINGLE LEPTOQUARK PRODUCTION AT TeV ENERGY $\gamma p$
      COLLIDERS}}}
 \author{ {\small $T.M.ALIEV^{1,2}
\,\,,D. A. DEMIR^{1} \,\,,
 E.ILTAN^{1} and \,\,N.K.PAK^{1}$}
 \\ {\it {\small 1) Physics Department, Middle East Technical University,
Ankara,Turkey}}
 \\ {\it {\small  2) International Centre of Theoretical
Physics,Trieste,Italy}}}
 \begin{titlepage}
 \maketitle
 \thispagestyle{empty}
 \begin{abstract}
 \baselineskip  .7cm
 The resolved and direct photon contributions to the single leptoquark
 (L) production process $\gamma p\rightarrow L e$ are analysed for both
 scalar (S) and vector (V) leptoquarks in detail. It is shown that
 resolved photon contribution dominates for $M_{L}\leq 300 GeV$.
 For $M_{V}\geq 1 TeV$ and $M_{S}\geq 0.5 TeV$ cross section
 is completely determined by the direct photon conribution. The
 vector leptoquarks are discussed for both gauge- and non-gauge
 cases seperately.
 \end{abstract}
 \end{titlepage}
 \baselineskip  .7cm
  \newpage
 \section{Introduction}
The theories beyond the Standard Model (SM) such as composite models[1],
grand unified theories[2], and $E_{6}$ superstring-inspired
models[3] predict the existence of leptoquarks carrying baryon and
lepton numbers simultaneously and having the electric charges $\pm 5/3;
\pm 4/3; \pm 2/3 $ and $\pm 1/3$ [4].

The production and possibility of the detection of leptoquarks have been
analysed in detail for, for instance, ep [5-7], hadronic [8], and
$e^{+}e^{-}$ colliders [4,9]. It is well known that high enery ep colliders
can be converted into a high energy $\gamma p$ collider with the help of
backscattered laser beams [10]. The single and double leptoquark production
in $\gamma p$ colliders are also analysed in the literature [11,12] without
taking the hadronic structure of photon into account. Namely they neglected
the resolved photon contribution. Furthermore in these works
the distribution of quarks and gluons in the proton are described by a
$Q^{2}$ independent parametrisation which may be misleading since the
c.m. energy for each subprocess supporting $\gamma p\rightarrow L e$
is not identical. In this work we shall analyse the production of
vector $(L=V)$ and scalar leptoquarks $(L=S)$ in $\gamma p$ colliders
by considering the resolved photon contribution as well.

The article is organized as follows: Section 2 describes the theoretical
basis and Section 3 is devoted to the numerical analysis and discussions.

\section{Total Cross Section for $\gamma p\rightarrow L e$ Scattering}
The production of leptoquarks in $\gamma p$ collisions can occur either
by direct or resolved photon processes. In the latter case photon interacts
with the proton through its hadronic components.It is clear that the
following processes are responsible for the $\gamma p\rightarrow L e$
scattering:
\begin{eqnarray}
\gamma q&&\rightarrow Le\nonumber\\
\gamma g&&\rightarrow Le\\
{\gamma}_{q} g &&\rightarrow Le\nonumber\\
{\gamma}_{g} q &&\rightarrow Le
\end{eqnarray}
where processes in (1) defines the direct photon scattering and those in
(2) defines resolved photon scattering.

The complete $SU(3)_{c}\times SU(2)_{L}\times U(1)_{Y}$
invariant Lagrangian in the low energy range ($M_{L}\approx 1 TeV$),conserving
baryon (B) and lepton (L) numbers, is given by [4]
\begin{eqnarray}
{\it{L}}= {\it{L}}^{f}_{F=2}+{\it{L}}^{f}_{F=0}+{\it{L}}^{scalar}+
{\it{L}}^{vector}
\end{eqnarray}
where $F=\mid 3B+L\mid$. The first two terms in the lagrangian are given by
the following expressions

\begin{eqnarray}
{\it{L}}^{f}_{F=2}&=&g_{1}(\bar{q}^{c}_{L}i\tau_{2} {\it{l}}_{L}+
                    \bar{u}^{c}_{R}e_{R})S_{1}\nonumber\\
&+&\tilde{g}_{1}\bar{d}^{c}_{R}e_{R}\tilde{S}_{1}+
g_{3}\bar{q}^{c}_{L}i\tau_{2}\vec{\tau}{\it{l}}_{L}.\vec{S}_{3}\nonumber\\
&+&g_{2}(\bar{d}^{c}_{R}\gamma^{\mu}{\it{l}}_{L}+
\bar{d}^{c}_{L}\gamma^{\mu}{\it{e}}_{R})V_{2\mu}\nonumber\\
&+&\tilde{g}_{2}\bar{u}^{c}_{R}\gamma^{\mu}{\it{l}}_{L}\tilde{V}_{2\mu}
+h.c.
\end{eqnarray}

\begin{eqnarray}
{\it{L}}^{f}_{F=0}&=&h_{2}(\bar{q}_{L}i\tau_{2} e_{R}+
                    \bar{u}_{R}{\it{l}}_{L})R_{2}\nonumber\\
&+&\tilde{h}_{2}\bar{d}_{R}{\it{l}}_{L}\tilde{R}_{2}+
h_{3}\bar{q}_{L}\vec{\tau}{\it{l}}_{L}.\vec{U}_{3}\nonumber\\
&+&h_{1}(\bar{q}_{L}\gamma^{\mu}{\it{l}}_{L}+
\bar{d}_{R}\gamma^{\mu}e_{R})U_{1\mu}\nonumber\\
&+&\tilde{h}_{1}\bar{u}_{R}\gamma^{\mu}e_{R}\tilde{U}_{1\mu}
+h.c.
\end{eqnarray}
In ${\it{L}}^{f}_{F=0}$ and ${\it{L}}^{f}_{F=2}$ there exist various
kinds of leptoquarks interacting with leptons and quarks. Here
$q_{L}$ and ${\it{l}}_{L}$ are $SU(2)_{L}$ left handed quark and
lepton doublets, $\psi^{c}=C\bar{psi}^{T}$ is the charge conjugated fermion
field. Among vector leptoquarks, $U_{1}, \tilde{U}_{1}$
are $SU(2)_{L}$  singlets, $V_{2}, \tilde{V}_{2}$ are left handed
$SU(2)_{L}$ doublets, and $\vec{U}_{3}$ is $SU(2)_{L}$ triplet.
Among scalar leptoquarks, $S_{1}, \tilde{S}_{1}$
are $SU(2)_{L}$  singlets, $R_{2}, \tilde{R}_{2}$ are left handed
$SU(2)_{L}$ doublets, and $\vec{S}_{3}$ is $SU(2)_{L}$ triplet.
Note that there are certain constraints on the leptoquark masses and
coupling constants from low energy experiments, which follows, for instance,
from the absence of $FCNC$ at the tree level, and from $D^{0}-\bar{D^{0}}$
and $B^{0}-\bar{B^{0}}$ mixings [13-16].

${\it{L}}^{scalar}$ in the Lagrangian describes the interaction of scalar
leptoquarks with the neutral gauge bosons, and is given by
\begin{eqnarray}
{\it{L}}^{scalar}=\sum_{i}[(D_{\mu}S^{i})^{\dagger}(D^{\mu}S^{i})
-M^{2}_{S}S^{\dagger}S]
\end{eqnarray}
where $S^{i}$ is the i-th scalar leptoquark field and
$D_{\mu}=\partial_{\mu}-ieA_{\mu}-ig_{s}\frac{\lambda^{a}}{2}A^{a}_{\mu}$
is the covariant derivative.

Unlike the scalar leptoquark case, $VV\gamma$ and $VVg$ vertices involve
an ambiguity depending on the nature of vector leptoquarks. For example,
if the vector leptoquarks are gauge bosons of an extended gauge group, then
the trilinear vertices are completely and unambigiously fixed by the gauge
invariance. When they are not gauge bosons, then the vector leptoquark
lagrangian can be regarded as an effective theory. It is clear that in
this effective theory there are many free parameters. In order to restrict
the number of these parameters, in addition to $SU(3)_{c}\times SU(2)_{L}\times
U(1)_{Y}$
invariance and B and L conservations, we impose the $CP$ invariance in the
effective lagrangian. Moreover we restrict ourselves to operators of
dimensionality 4 or less. There is only one operator
$ig\kappa V^{\mu}V_{\mu\nu}V^{\nu}$ complying with these conditions  and it
describes the anomalous magnetic moment contribution to the trilinear vertex.

Thus the effective gauge boson- vector leptoquark lagrangian
${\it{L}}^{vector}$ conserving $CP$ and containing operators of dimension
4 or less is given by (see, for example, [17])
\begin{eqnarray}
{\it{L}}^{vector}&=&\sum_{i}\{-\frac{1}{2}V^{\dagger}_{i\mu\nu}V^{i\mu\nu}
+M^{2}_{V}V^{\dagger}_{i\mu}V^{i\mu}\nonumber\\
&-&i\sum_{j=\gamma,g}g_{j}\kappa_{j}V_{i\mu}G^{j\mu\nu}V_{i\nu}\}
\end{eqnarray}
where $i$ runs over all vector leptoquarks,
$V_{\mu\nu}=D_{\mu}V_{\nu}-D_{\nu}V_{\mu}$ is the leptoquark field strength
and $G_{j\mu\nu}$ is the field strength tensor of photon ($j=\gamma$)
or gluon ($j=g$). In what follows we shall take anomalous couplings of
photon and gluon identical; $\kappa_{\gamma}=\kappa_{g}=\kappa$.

One can readily obtain  the Feynman rules for trilinear vertices
immediately from ${\it{L}}^{vector}$
\begin{eqnarray}
VV\{\begin{array}{c}\gamma\\
g\end{array}\}&=&\{\begin{array}{c}ieQ\\ig_{s}\frac{\lambda^{a}}{2}\end{array}\}\{
(k_{2}-k_{3})_{\mu}g_{\alpha\beta}+(k_{3}-\kappa
k_{1})_{\alpha}g_{\mu\beta}\nonumber\\
&+&(k_{1}(1+\kappa)-k_{2})_{\beta}g_{\mu\beta}\}
\end{eqnarray}
where all momenta are incoming. Here $k_{1},k_{2}$ and $k_{3}$ are the 4-
momenta of photon (gluon) and leptoquarks respectively. We will next
turn our attention to the calculation of the cross section for
$\gamma p\rightarrow L e$ scattering.

The total cross section for $\gamma p\rightarrow L e$ scattering has the form
\begin{eqnarray}
\sigma(\gamma p\rightarrow L e)=\sigma_{dir}(\gamma p\rightarrow L e)
                               +\sigma_{res}(\gamma p\rightarrow L e)
\end{eqnarray}
where $\sigma_{dir}$ and $\sigma_{res}$ represent the contributions of the
direct and resolved photons respectively. The total cross section for
the leptoquark production by direct photon interaction can be obtained
by folding the cross section for the elementary process (1) with the
the photon distribution in electron  and quark distribution in the
proton:
\begin{eqnarray}
\sigma_{dir}(\gamma p\rightarrow L e)&=&\int_{\lambda}^{0.83} dx
\int_{\frac{\lambda}{x}}^{1} dy\,
                   f_{\gamma/e}(x)\,f_{q/p}(y,\frac{xys}{2})\,\sigma_{\gamma
q}(xys)
\end{eqnarray}
In the same manner, the total cross section for leptoquark production by
resolved photon contribution can be obtained by folding the cross section
of the elementary process (2) with quark (gluon) distribution in the
photon, gluon (quark) distribution in the proton and photon distribution
in the electron:

\begin{eqnarray}
\sigma_{res}(\gamma p\rightarrow L e)&=&
\int_{\lambda}^{0.83} dx \int_{\frac{\lambda}{x}}^{1} dy
\int_{\frac{\lambda}{xy}}^{1} dz
f_{\gamma/e}(x)[f_{g/\gamma}(y,
\frac{xys}{2})f_{q/p}(z,\frac{xyzs}{2})\nonumber\\&+&
f_{q/\gamma}(y,\frac{xys}{2})f_{g/p}(z,\frac{xyzs}{2})]\;\sigma_{g q}(xyzs)
\end{eqnarray}
In (10) and (11) $f_{a/b}(x,Q^{2})$ is the $Q^{2}$ dependent distribution
function of the parton $a$ in the hadron $b$ ($f_{\gamma/e}(x)$ is an
exception).
In all these functions we have set $Q^{2}=\frac{\hat{s}}{2}$ where $\hat{s}$ is
the invariant
mass flow to the subprocess under concern, $\sqrt{s}$ is the c.m. energy
of the collider and $\lambda=(M_{L}+m_{q})^{2}/s$, where $M_{L}$ and $m_{q}$
are the leptoquark and quark masses respectively.

Let us now discuss the construction of the formulae (10) and (11)
in the case of ${\gamma}_{g}q\rightarrow Le$, as an example. Here $s$
is the c.m. energy squared of $ep$ collider. Only fraction $x$ of $s$ enters
the $\gamma p$ system, so $s_{\gamma p}=xs_{ep},\;\;0\leq x\leq 1$.
Now hadronic components of photon mediate some fraction $y$ of
$s_{\gamma p}$, so $s_{gp}=ys_{\gamma p}, \;\;0\leq y\leq 1$.
Similarly, quark coming off the proton takes away some fraction $z$ of
$s_{gp}$ so $s_{gq}=zs_{gp},\;\;0\leq z\leq 1$.

The total cross section of the elementary subprocess (1) is given by
\begin{eqnarray}
\sigma^{V}_{\gamma q}&=&A_{\gamma}\{
\ln{(\frac{a - 1}{b})}[4q^{2}_{2}-8a(a-1)q_{2}(q_{3}-q_{1})]\\
&+&\ln{a}[4q_{3}q_{1}(\kappa+1)+q^{2}_{3}(\kappa^{2}+6\kappa+1)+
+8aq_{3}(q_{3}+a(q_{1}-q_{2}))\nonumber\\
&+&\frac{q_{3}(\kappa-1)}{a}\{2(q_{2}+q_{1})-q_{3}(\kappa+1))\}]\nonumber\\
&+&8a^{2}\{q_{2}q_{1}-8q_{3}(q_{3}+q_{1})+2q^{2}_{1}\}
+a\{q^{2}_{2}-q_{2}q_{3}(\kappa+1)-6q_{2}q_{1}\nonumber\\&+&(1/4)q^{2}_{3}\kappa(\kappa+2)
+(33/4)q^{2}_{3}-q_{3}q_{1}(-5\kappa+3)-3q^{2}_{1}\}\nonumber\\
&+&\frac{1}{a}\{q^{2}_{2}+q_{2}q_{3}(\kappa-3)+2q_{2}q_{1}+
(1/4)q^{2}_{3}(\kappa^{2}+26\kappa+9)\nonumber\\
&+&q_{3}q_{1}(\kappa-3)+q^{2}_{1}\}
-2q^{2}_{2}+4q_{2}q_{3}-4q_{2}q_{1}-(1/2)q^{2}_{3}(\kappa^{2}+14\kappa+5)\nonumber\\
&+&4q_{3}q_{1}(\kappa+2)\}\nonumber
\end{eqnarray}
and
\begin{eqnarray}
\sigma^{S}_{\gamma q}&=&A_{\gamma}\{
 \ln{(\frac{1- a}{b})}[2a(a-1)q_{2}(q_{1}-q_{3})+q_{2}^2]\\
&+&\ln{a}[2aq_{3}\{q_{3}+a(q_{1}+q_{2})\}]+
a^{2}\{(1/2)q_{1}^2-2q_{1}q_{2}\nonumber\\&-&2q_{1}q_{3}-2q_{3}^{2}\}
+a\{-q_{1}^2+4q_{1}q_{2}+2q_{1}q_{3}+2q_{3}^2\}\nonumber\\
&+&(1/2)q_{1}^{2}-2q_{1}q_{2}\}\nonumber
\end{eqnarray}
where superscripts $V$ and $S$ refer to the vector and scalar leptoquark
productions respectively. In (4) and (5) $a=M^{2}_{L}/s$, $b=m^{2}_{q}/s$,
$q_{1}=q$, $q_{2}=1$, $q_{3}=q+1$ ($q$ quark charge), and $\kappa$ is
the anomalous coupling of leptoquarks to photon and gluon [17]. The factor
$A_{\gamma}$ is given by
\begin{eqnarray}
A_{\gamma}&=&\frac{\pi\alpha^{2}}{2s}.
\end{eqnarray}

The crossections $\sigma^{V}_{g q}$ and $\sigma^{S}_{g q}$ could be obtained
from (4) and (5) with the following replacements:
\begin{eqnarray}
\sigma^{V,S}_{g q}=\sigma^{V,S}_{\gamma q}(q_{1}=1,q_{2}=0,q_{3}=1,
A_{\gamma}\rightarrow A_{g})
\end{eqnarray}
where
\begin{eqnarray}
A_{g}&=&\frac{\pi\alpha\alpha_{s}}{12s}.
\end{eqnarray}
and $\alpha_{s}$ is given by
\begin{eqnarray}
\alpha_{s}(Q^{2})=\frac{12\pi}{(33-2f)\ln{Q^{2}/\Lambda^{2}}}
\end{eqnarray}
up to one-loop accuracy.

Now, let us consider the large $s$ behaviour of the cross section for
the subprocess in (1) for the arbitrary values of $\kappa$. From (12)
one can easily obtain
\begin{eqnarray}
\sigma^{V}_{\gamma q}=\frac{\pi \alpha^{2}(q+1)^{2}}{M_{V}^{2}}
[\frac{1}{2}\ln{\frac{s}{m^{2}_{V}}}(\kappa-1)^{2}+\frac{1}{8}(\kappa^{2}
+30\kappa+1)]
\end{eqnarray}
in the limit $s>>M^{2}_{V}$. We see that for $\kappa \neq 1$ the
cross section grows logarithmically.  This is due to the $t$- channel
contribution. As we noted before, if $V$ is a non- gauge particle ($\kappa \neq
1$)
this logarithmic dependence can be considered as the low energy
manifestation
of a more fundamental theory at a higher energy scale. So, according to
the effective theory description, the behaviour of the cross section is
acceptable as long as energy is sufficiently low. At high energies, the
effective theory is superseeded by a more fundamental theory, where the
increase of the cross section with $s$ is stopped  and unitarity is
preserved. If $V$ has the gauge nature (where $\kappa=1$), the cross
section reaches the constant value
\begin{eqnarray}
\sigma^{V}_{\gamma q}=\frac{4\pi \alpha^{2} (q+1)^{2}}{M^{2}_{V}}
\end{eqnarray}
which is similar to the single $W$ boson production in the reaction
$\gamma p\rightarrow W X$ [17] in the standard model.

After giving the expression for subprocess cross sections we now
turn to the expicit expressions for the distribution functions
in (10) and (11).

The function $f_{\gamma/e}(x,Q^{2})$ is the energy spectrum of the
backscattered
laser photons [10]
\begin{eqnarray}
f_{e/\gamma}(x)=\frac{1}{D(\zeta)}[1-x +\frac{1}{1-x}-\frac{4x}{\zeta(1-x)}
+\frac{4 x^{2}}{\zeta^{2}(1-x)^{2}}]
\end{eqnarray}
where
\begin{eqnarray}
D(\zeta)=(1-\frac{4}{\zeta}-\frac{8}{\zeta^{2}})\ln{(1-\zeta)}
         \frac{1}{2}+\frac{8}{\zeta}-\frac{1}{2(1+\zeta)^{2}}
\end{eqnarray}
with $\zeta=4.82$. The maximum value of $x$ is found as
$x_{max}=\frac{\zeta}{\zeta+1}=0.83$ which is the upper limit of
the $x$ integral in (2) and (3).

In describing the quark and gluon distributions in the proton
we shall use the results of [18] where a $Q^{2}$ dependent
parametrisation is given. We shall not reproduce all the details
of the parametrisations here, instead we summarize the
general form of the functions and refer the reader to the
references for details. $f_{q/p}(x,Q^{2})$ parametrizes
the quark (sea plus valence) distributions in the proton
\begin{eqnarray}
xf_{q/p}(x,Q^{2})&=&N\frac{A_{2}}{{\it{B}}(A_{3}+1,A_{1}/A_{2})}
x^{A_{1}}(1-x^{A_{2}})^{A_{3}}\\
&+&A_{4}(1+A_{5}x+A_{6}x^{2})(1-x)^{A_{7}}+A_{8}e^{-A_{9}x}
\end{eqnarray}
where ${\it{B}}(x,y)$ is the Euler's Beta function, $N$ equals 2
for $u$ quark and 1 for $d$ quark. The coefficients $A_{i}$ (i=1,..,9)
are tabulated in [18] and they are explicit functions of
\begin{eqnarray}
\bar{s}=\ln{\{ \frac{\ln{(Q^{2}/\Lambda^{2})}}{\ln{(Q_{0}^{2}/\Lambda^{2})}}
\}}
\end{eqnarray}
where $Q^{2}_{0}=4 GeV^{2}$ and $\Lambda=0.4 GeV$.

The gluon distribution in the proton is parametrised by $f_{g/p}$
having the expression [18]
\begin{eqnarray}
xf_{g/p}(x,Q^{2})&=&B_{1}(1+B_{2}x+B_{3}x^{2})(1-x)^{B_{4}}+B_{5}e^{-B_{6}x}
\end{eqnarray}
and again the coefficients $B_{i}$ are functions of $\bar{s}$ and are
tabulated in [18].

For the gluon and quark distributions in the photon we shall use the
$Q^{2}$ dependent parametrisation given in [19]. The quark distribution
in the photon is parametrised by $f_{g/\gamma}(x,Q^{2})$ which is given by
\begin{eqnarray}
xf_{g/\gamma}(x,Q^{2})=C_{1}x^{C_{2}}(1-x)^{C_{3}}
\end{eqnarray}
where the coefficients $C_{i}$ are functions of
\begin{eqnarray}
t=\ln{(Q_{0}^{2}/\Lambda^{2})}e^{\bar{s}}
\end{eqnarray}
and are tabulated in [19].

The quark distribution (sea plus valence) in the photon is parametrised
by
\begin{eqnarray}
f_{q/\gamma}(x,Q^{2})=A_{f}q_{v}(x,Q^{2})+B_{f}q_{s}(x,Q^{2})
\end{eqnarray}
where the coefficients $A_{f}$ and $B_{f}$ [19] change as the number of
flavours $f$ changes ( in connection with the momentum scale $Q^{2}$) and
the functions $q_{v}(x,Q^{2})$ and $q_{s}(x,Q^{2})$ ( $q^{\gamma}_{NS}$ and
$\Sigma^{\gamma}$ respectively, in the notation of [19]) are given by
\begin{eqnarray}
xq_{j}(x,Q^{2})&=&x\frac{x^{2}+(1-x)^{2}}{D_{1j}-D_{2j}ln(1-x)}
                 +D_{3j}x^{D_{4j}}(1-x)^{D_{5j}}
\end{eqnarray}
where $j=v,s$, and the coefficients $D_{ij}$ $(i=1..5, j=v,s)$ are
given in [19].

\section{Numerical Analysis}
We will now analyze the total cross section $\sigma(\gamma p\rightarrow L e)$
defined in (9) for vector and scalar leptoquarks. We based our analysis
only to the first generation so that the quark entering the scattering
process is either $u$ or $d$, producing in the final state leptoquarks
of electromagnetic charge 5/3 or 2/3 [4] respectively. Although there
are many accelerators [11,12] deserving analysis under such a work, for
our purpose it is sufficient to analyze a single accelerator which we
choose to be $LHC+TESLA$ with $\sqrt{s}=5.5 TeV$.

Fig.1 and Fig.2 show the total cross section in (9), at $\sqrt{s}=5.5 TeV$, for
$V_{5/3}$ (gauge particle) and $S_{5/3}$ respectively. We give Fig.1 and Fig.2
to demonstrate
the relative magnitude of the direct and resolved photon contributions
as the leptoquark mass changes. These two are typical examples applicable to
all other cases. From Fig. 1 and Fig. 2 we see that for $M_{L}\leq 300 GeV$
the total cross section in (9) is strongly dominated by resolved photon
contribution in (11). Moreover, Fig. 1 and Fig. 2 show, respectively,  that
for $M_{V}\geq 1 TeV$ and $M_{S}\geq 0.5 TeV$ the total cross section
in (9) is completely determined by the direct photon contribution in (10).
We see that the leptoquark mass range where the resolved photon contribution
is non negligible is within the mass bounds given in [17]; thus, the effects
of the hadronic component of photon in $\gamma p$ colliders are directly
observable in future experiments.

Fig. 3 shows the dependence of the total cross section in (12) on $M_{V}$
for different values of $\kappa$. From this figure we see that for non-
gauge $V$, the cross section in (12) is considerably enhanced (suppressed) as
$\kappa$ grows (falls) to higher (lower) values from unity. Especially
the $\kappa \geq 1$ case is interesting, because enhancement in the
cross section is large for low values of $M_{V}$ where the resoved photon
contribution dominates. Lastly, we observe from this figure that the cross
section for $V_{5/3}$ is always larger than that for $V_{2/3}$.

Finally, in Fig.4 we show the variation of the total cross section in (13)
with $M_{S}$. We conclude from this figure that the total cross section
for $S_{5/3}$ is always larger than that for $S_{2/3}$ ( see also [11,12]).

In conclusion, we have discussed the single leptoquark production
in $\gamma p$ colliders for scalar and vector leptoquarks. We
have analysed the contribution of the hadronic component of the
photon to the total cross section. Moreover, gauge- and non- gauge-
vector leptoquarks are discussed seperately in terms of their contribution
to the total cross section.

Acknowledgement: One of the authors (T.M.A.) sincerely thanks
International Centre of Theoretical Physics, Trieste, and Prof.
S. Randjbar-Daemi for hospitality, where part of this work is done.

\newpage
\begin{description}
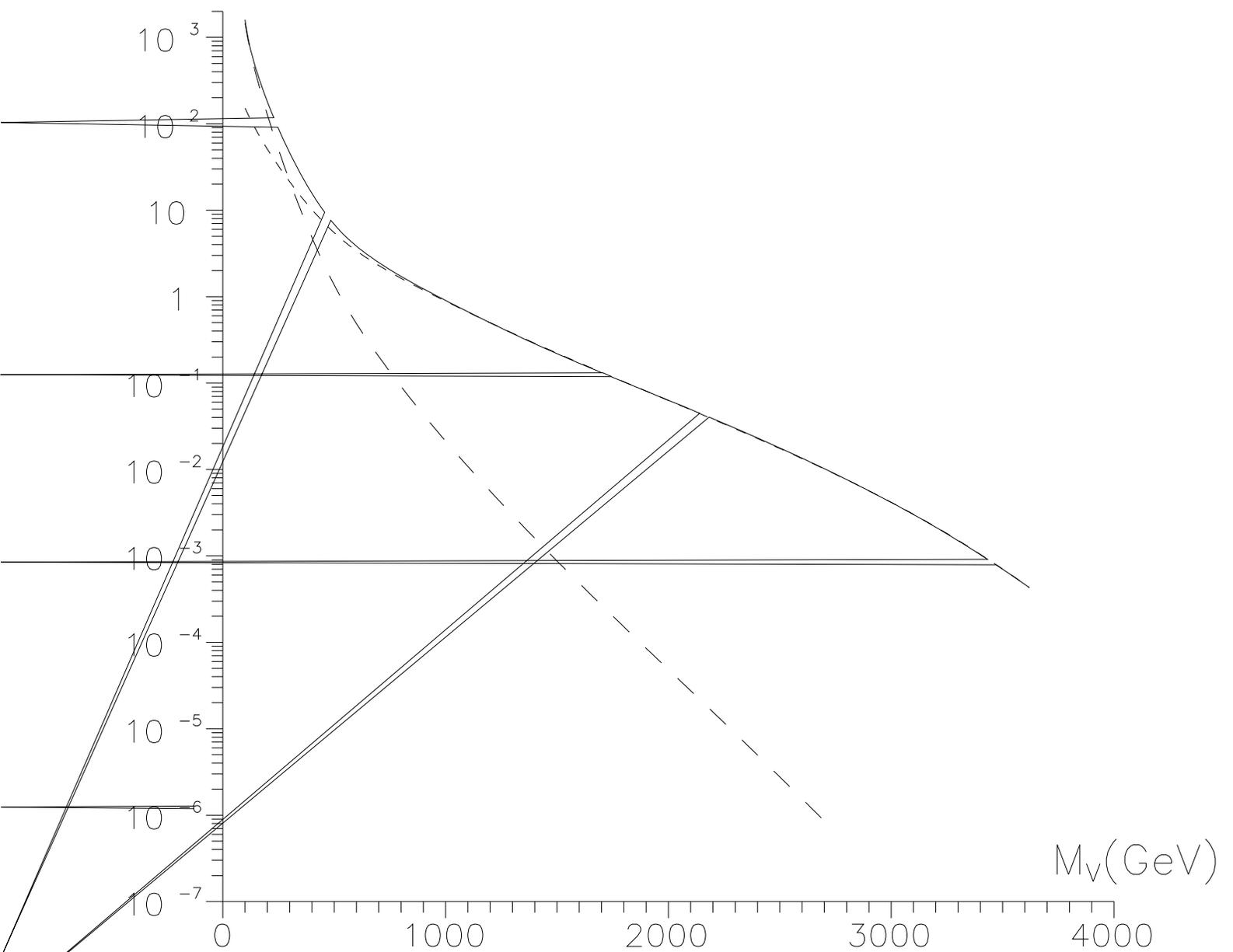
\item[{\bf Figure 1}:] For $V_{5/3}$ (gauge particle) at $\sqrt{s}=5.5 TeV$,
contributions of resolved photon (dashed) and direct photon (short- dashed)
to the total crossection (solid).
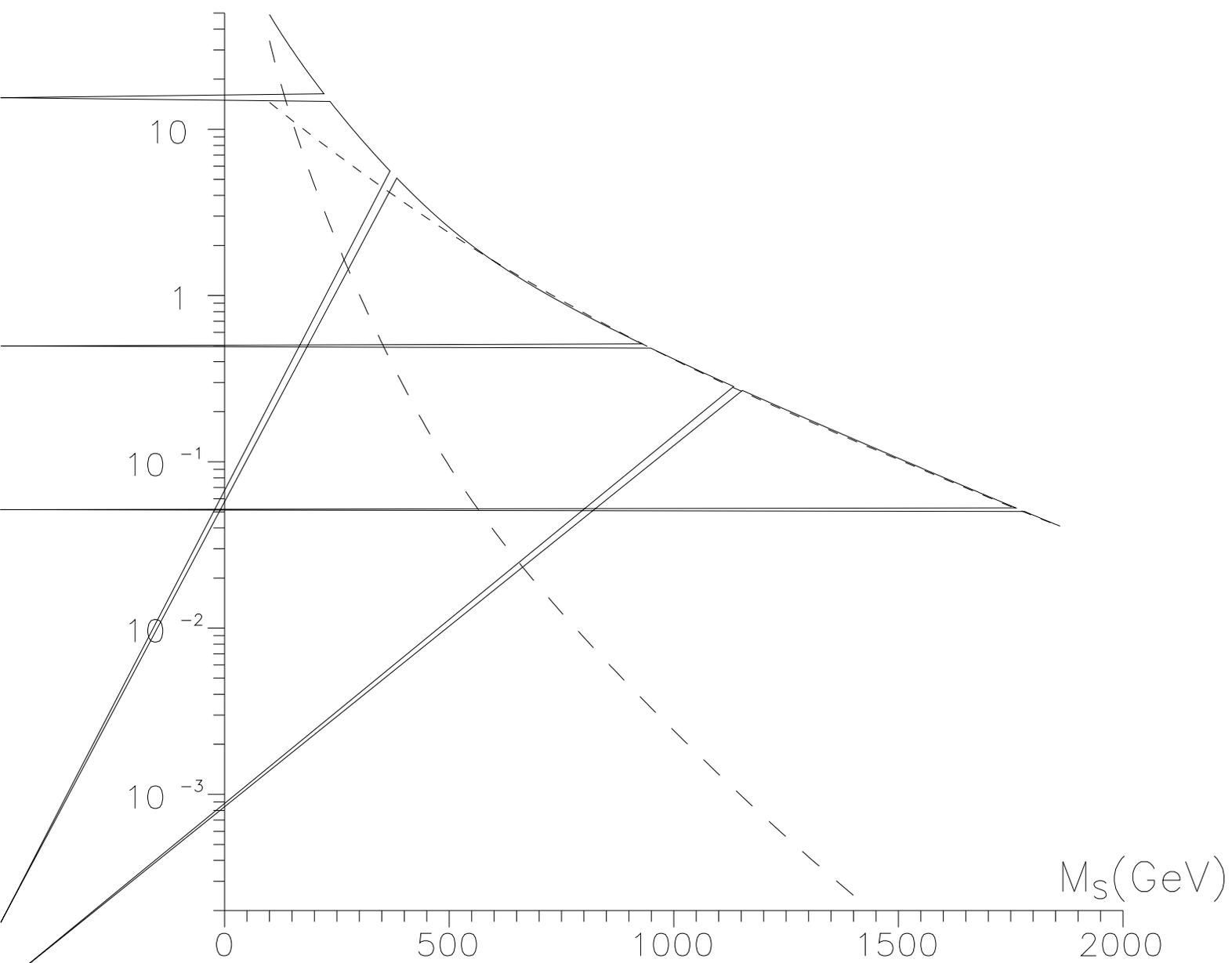
\item[{\bf Figure 2}:] The same as in Fig. 1 but for $S_{5/3}$
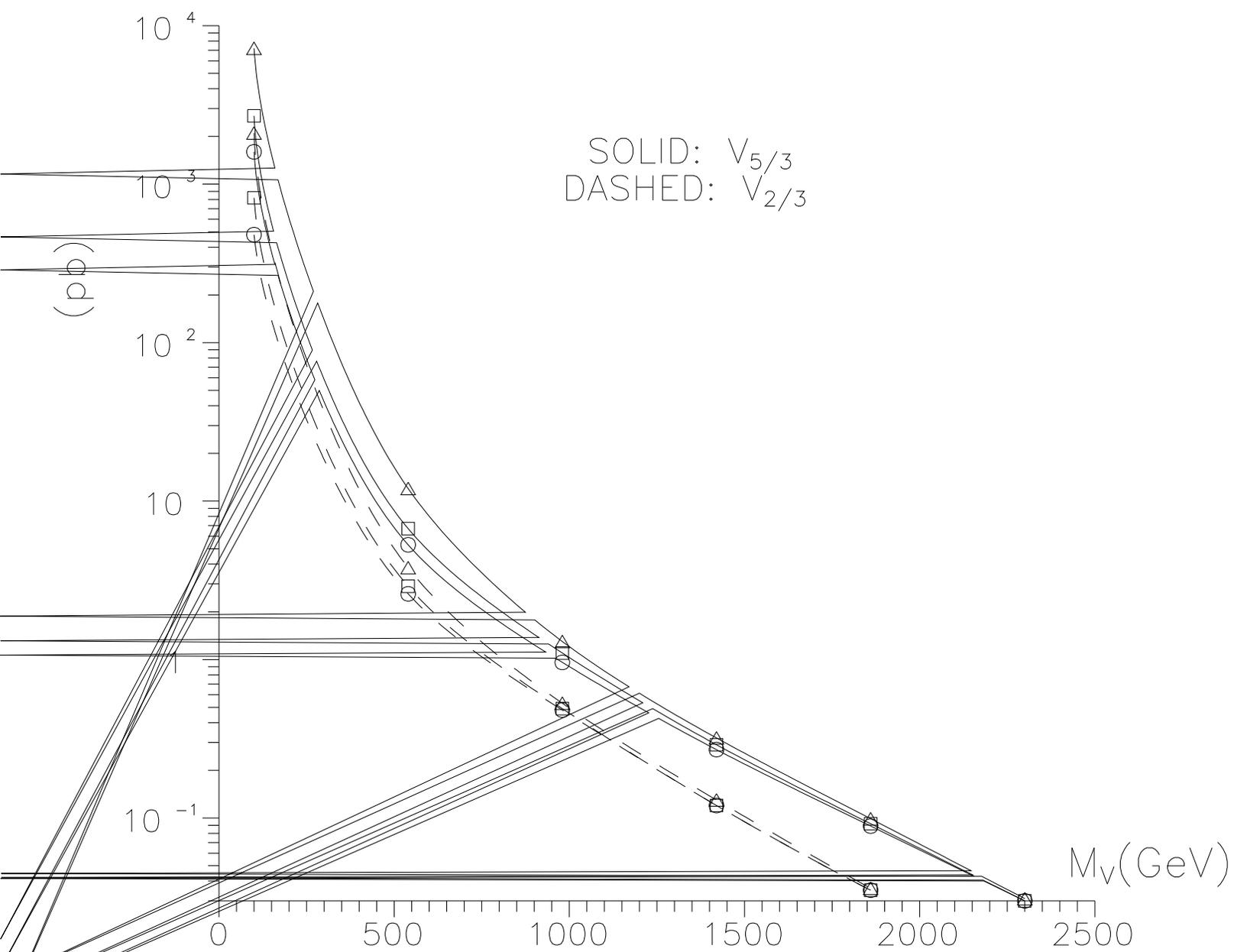
\item[{\bf Figure 3}:] Variation of the total crossection for
vector leptoquarks as a function of leptoquark mass for different values
of anomalous coupling. Here circle, square and triangle corresponds to
$\kappa=0.5$, $\kappa=1.0$ and $\kappa=2.0$ respectively.
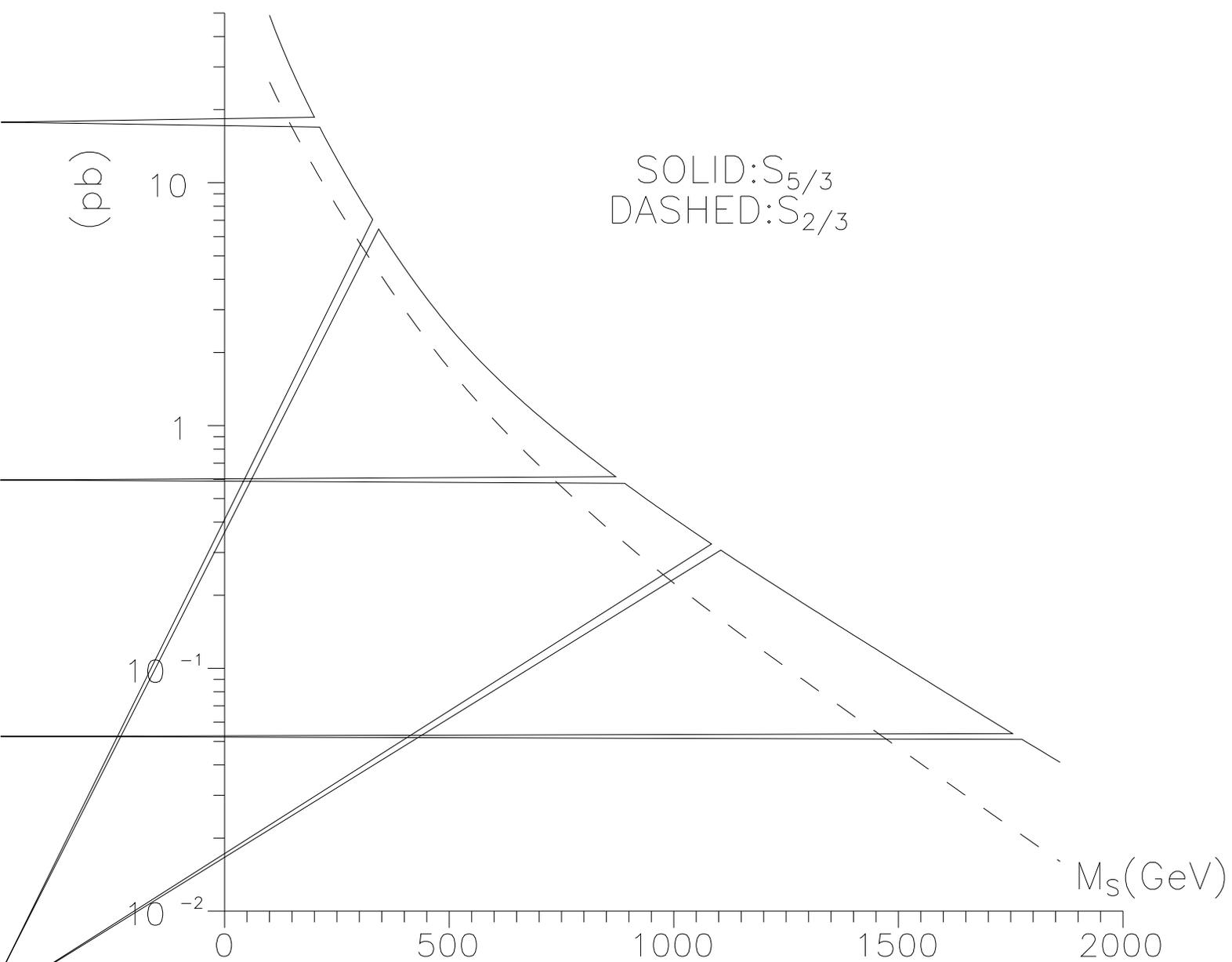
\item[{\bf Figure 4}:] Variation of the total crossection for
scalar leptoquarks as a function of leptoquark mass.

\end{description}

\newpage
\begin{thebibliography}{99}
\bibitem{R1} L. Abbott and E.Farhi, Phys. Lett. B101(1981)69;
             Nucl.Phys. B189(1981)547.
\bibitem{R2} P. Langacker, Phys. Rep. 72(1981)193.
\bibitem{R3} J. L. Hewett and T. Rizzo, Phys. Rep. 183(1989)193.
\bibitem{R4} J. Blumlein and R. Ruckl, Phys. Lett. B304(1993)337.
\bibitem{R5} J. Wudka, Phys. Lett. B167(1986)337.
\bibitem{R6} J.F. Gunion and E. Ma, Phys. Lett. B195(1987)257.
\bibitem{R7} W. Buchmuller et. al. Phys. Lett. B191(1987)442.
\bibitem{R8} O.J.P. Eboli and A.V. Olinto, Phys. Rev. D38(1988)3461.
\bibitem{R9} J. L. Hewett and S. Pakvasa, Phys. Lett. B227(1989)178.\\
             O.J.P. Eboli et. al., Phys. Lett. B311(1993)147.
\bibitem{R10} I. F. Ginzburg et.al., Nucl. Ins. Methods, 5(1984)219.
\bibitem{R11} S. Atag et.al., Phys. Lett. B326(1994)185.
\bibitem{R12} S. Atag et.al., J. Phys. G: Nucl. Part. Phys 21(1995)1189.
\bibitem{R13} W. Buchmuller et. al., Phys. Lett. B177(1987)377.
\bibitem{R14} O. Shanker, Nucl. Phys. B206(1982)49.
\bibitem{R15} M. Leurer, Phys. Rev. D50(1994)536.
\bibitem{R16} S. Davidson et. al., Z. Phys. C61(1994)613.
\bibitem{R17} J. Blumlein and E. Boos, Prep. DESY 94-144 (1994).
\bibitem{R18} M. Gluck, E. Hoffmann and E. Reya, Z.Physik C13(1982)119.
\bibitem{R19} M. Drees and K. Grassie, Z.Physik C28(1985)451.
\end{thebibliography}
\end{document}